\documentclass[twocolumn,prb,showpacs]{revtex4}
\usepackage{amsmath,graphicx}
\begin{document}
\title{The Keldysh action of a multi-terminal time-dependent scatterer.}
\author{I. Snyman}
\affiliation{Instituut-Lorentz, Universiteit Leiden, P.O. Box 9506, 2300 RA Leiden, The Netherlands}
\author{Y.V. Nazarov}
\affiliation{Kavli Institute of Nanoscience, Delft University of Technology, 2628 CJ Delft, The Netherlands}
\date{\today}
\begin{abstract}
We present a derivation of the Keldysh action of a general multi-channel
time-dependent scatterer in the context of the Landauer-B\"uttiker approach. 
The action is a convenient building block in the theory of quantum transport. 
This action is shown to take a compact form that only involves the scattering
matrix and reservoir Green functions. We derive two special cases of the general result,
one valid when reservoirs are characterized by well-defined filling factors, the
other when the scatterer connects two reservoirs. 
We illustrate its use by considering 
Full Counting Statistics and the Fermi Edge Singularity.
\end{abstract}
\maketitle 
\section{Introduction}
The pioneering work of Landauer\cite{Lan57,But85} and B\"uttiker\cite{But90,But92} 
lay the foundations for what is now known
as the scattering approach to electron transport. The basic tenet is
that a coherent conductor is characterized by its scattering matrix. More precisely
the transmission matrix defines  a set of transparencies for the various channels
or modes in which the electrons propagate through the conductor. 
As a consequence, conductance 
is the sum over transmission probabilities. Subsequently, it was discovered that 
the same transmission probabilities fully determine the current noise, 
also outside equilibrium, where the fluctuation-dissipation theorem does not hold\cite{Bla00}. 

Indeed, as the theory of Full Counting Statistics \cite{Lev93,Lev96,Kli03} later revealed, the complete probability
distribution for outcomes of a current measurement is entirely characterized by the transmission
probabilities of the conductor. The fact that the scattering formalism gives such an elegant and 
complete description inspired some to revisit established results. 
Thus for instance interacting problems such as the Fermi Edge Singularity \cite{Mah67,Noz69}
was recast in the language of the scattering approach\cite{Mat92,Yam82,Aba04,Aba05}. 
The scattering approach has 
further been employed successfully in problems where a coherent conductor interacts with 
other elements, including, but not restricted to, measuring devices and an electromagnetic
environment \cite{Kin03,Kin04,Tob06,Sny07}. It is also widely applied to study transport in
mesoscopic superconductors \cite{Bee92}.

Many of these more advanced applications are unified through a method developed by Feynman and Vernon 
for characterizing the effect of one quantum system on another when they are coupled \cite{Fey63}.
The work of Feynman and Vernon dealt with the effect of a bath of oscillators coupled to a quantum
system. It introduced the concept of a time-contour describing propagation first forwards then backwards
in time. By using the path-integral formalism, it was possible to characterize the bath by an 
``influence functional''
that did not depend on the system that the bath was coupled to. This functional was treated non-perturbatively.
A related development was due to Keldysh \cite{Kel64}. While being a perturbative diagrammatic technique, it allowed
for the treatment of general systems and shared the idea of a forward and backward time-contour with Feynman and Vernon. 
Applications involving the scattering approach require both
the notion of the non-perturbative influence functional and the generality of Keldysh's formalism. 
Until now, the combination of the Feynman-Vernon method with the scattering approach was done
on an case-specific basis. Only those elements relevant to the particular application under consideration
were developed. In this paper we unify previous developments by deriving general formulas for
the influence functional, or equivalently the Keldysh action of a general scatterer connected
to charge reservoirs.

The Keldysh action of a general scatterer can be considered as a building block. Its ``interface''
is the set of fields $\chi_\pm(t)$. Through this interface, the actions
of many conductors can be combined into quantum circuits. As in the case of classical electronics,
a simple set of rules, applied at the nodes of such a circuit, suffice to describe the behavior
of the whole network \cite{Naz99,Naz94}. 

The influence functional $\mathcal Z[\chi_\pm]$ and the Keldysh action 
$\mathcal A[\chi_\pm]=\ln \mathcal Z[\chi_\pm]$ depend on two sets of time-dependent fields
$\chi_+(t)$ and $\chi_-(t)$ corresponding to forward and backward evolution in time with
different Hamiltonians. 
Recently, the  situation was considered where $\chi_\pm$ are time-independent but
the scatterer was allowed to fluctuate in time \cite{Aba07}. We consider the case where also
the fields $\chi_\pm$ are time-dependent.
Functional derivatives with respect to these fields generate 
cumulants of the distribution of outcomes for the measurement of the degrees of freedom coupled to 
$\chi_\pm$. Since these fields enter the Hamiltonian of the scatterer as a time-dependent potential
energy term, their effect is captured by the scattering matrix. Since the fields differ for
forward and backward evolution, the scattering matrices for forward and backward evolution
differ. 

Our main result is summarized by a formula for this Keldysh action.
\begin{equation}
\mathcal A[\hat s]=
{\rm Tr\;ln}\left[\frac{1+\hat G}{2}+\hat s\frac{1-\hat G}{2}\right]-{\rm Tr\;ln}\;\hat s_-.
\label{eq0}
\end{equation}
In this formula, $\hat G$ is the Keldysh Green function characterizing the reservoirs connected to the
scatterer \cite{Ram86}. It is to be viewed as an operator with kernel $G(\alpha,\alpha';c;t,t')\delta_{c,c'}$
where the Keldysh indices $\alpha,\,\alpha'\in\{+,-\}$ refer to time-contour ordering,
$c,\,c'\in Z$ refer to channel space, and $t,\,t'\in R$ are continuous time indices.
The dependence on the fields $\chi_\pm$ is carried by the time-dependent 
scattering matrix $\hat s$, that also has Keldysh structure, owing to forward and backward time-evolution
with different Hamiltonians. Explicitly, it is to be viewed as an operator with kernel
$s(\alpha;c,c';t)\delta_{\alpha,\alpha'}\delta(t-t')$ where indices carry the same meaning as in the
kernel of $\hat G$.  This formula is completely general. 
\begin{enumerate}
\item{It holds for multi-terminal devices with more than two reservoirs.} 
\item{It holds for devices such as Hall bars where particles in a single chiral channel enter
and leave the conductor at different reservoirs.}
\item{It holds when reservoirs cannot be characterized by stationary filling factors. Reservoirs
may be superconducting, or contain ``counting fields'' coupling them to a dynamical electromagnetic
environment or a measuring device.}
\end{enumerate}

When the reservoirs can indeed be characterized by filling factors $\hat f(\varepsilon)$, 
the Keldysh structure can explicitly be traced out to yield
\begin{equation}    
\mathcal A[\hat s_+, \hat s_-]={\rm Tr\;ln}\left[\hat s_-(1-\hat f)+\hat s_+\hat f\right]
-{\rm Tr\;ln}\;\hat{s}_-.\label{eqa}
\end{equation}
In this expression operators retain channel structure and time structure. The time-dependent
scattering matrices $\hat{s}_\pm$ have kernels $s_\pm(c,c';t)\delta(t-t')$ that 
depend on the field $\chi_\pm(t)$. In ``time'' representation,
$\hat f$ is the Fourier transform to time of the reservoir filling factors, and as such has a kernel
$f(c;t,t')\delta_{c,c'}$ diagonal in channel space and depending on two times.
This formula is of the same type as the Levitov-Lesovik formula for zero frequency Full Counting Statistics (FCS) 
\cite{Kli03}, but contains information about finite frequencies due to the arbitrary time-dependence of 
$\hat s_\pm$.

Another formula may be derived from Eq. (\ref{eq0}), valid for two terminal devices. 
Each terminal may still be connected to the scatterer by an arbitrary number of channels.
We denote
the two terminals left (L) and right (R).
In this case the reservoir Green function  has the form
\begin{equation}
\hat G=\left(\begin{array}{cc}\check G_L&0\\0&\check G_R\end{array}\right)_{\rm channel\,space}
\end{equation}
where $\check G_{L(R)}$ have no further channel space structure. 
Matrix structure in Keldysh and time indices (indicated by a check sign)
is now retained in the trace, but the channel structure is traced out. Thus is obtained
\begin{equation}
\mathcal A[\chi_\pm]=\frac{1}{2}\sum_n{\rm Tr}\,{\rm ln}\left[1+T_n
\frac{\left\{\check G_L[\chi_\pm],\check G_R[\chi_\pm]\right\}-2}{4}\right].\label{eqb}
\end{equation}
In this expression, the field dependence $\chi_\pm$ is shifted to the Keldysh Green functions
$\check G_L$ and $\check G_R$ of the left and right reservoirs.
This formula makes it explicit that the conductor is completely characterized by its transmission 
eigenvalues $T_n$. 

The plan of the text is as follows. After making the necessary definitions, we derive Eq. (\ref{eq0})
from a model Hamiltonian. The derivation makes use of contour ordered Green functions
and the Keldysh technique. Subsequently, we derive the special cases of Eq. (\ref{eqa}) and 
Eq. (\ref{eqb}). While the formulas (\ref{eqa}) and (\ref{eqb}) have appeared in the literature
before, as far as we know, there has not yet appeared a formal derivation. 

We conclude by applying the formulas to several generic set-ups, and verify 
that results agree with the existing literature. Particularly, we explain in detail how the present work is connected 
to the theory of Full Counting Statistics and to the scattering theory of the Fermi Edge Singularity.
  
\section{Derivation}
\begin{figure}[h]
\begin{center}
\includegraphics[width=.95\linewidth]{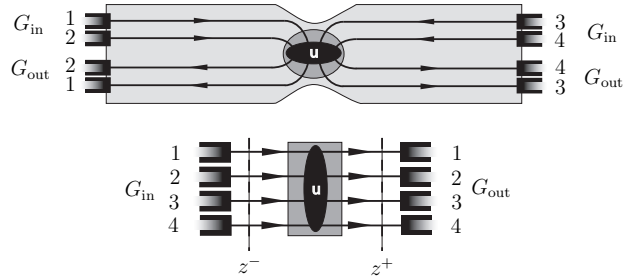}
\caption{We consider a general scatterer connected to reservoirs. The top figure is a diagram of one possible physical
realization of a scatterer. Channels carry electrons towards and away from a
scattering region (shaded dark gray) where inter-channel scattering takes place. Reservoirs
are characterized by Keldysh Green functions $G_{\rm in\,(out)}$. These Green functions also
carry a channel index, in order to account for, among other things, voltage biasing.
In setups such as the the Quantum Hall experiment where there is a Hall voltage, $G_{\rm in}$
will differ from $G_{\rm out}$, while in an ordinary QPC, the two will be identical.
The bottom figure shows how the physical setup is represented in our model. Channels are 
unfolded so that all electrons enter at $z^-$ and leave at $z^+$.}   
\end{center}
\end{figure}
We consider a general scatterer connecting a set of charge reservoirs. We allow the scatterer to be time-dependent. 
A sufficient theoretical
description is provided by set of transport channels
interrupted by a potential that causes inter-channel scattering. 
We consider the regime where the scattering matrix is energy-independent in the transport energy
window. Since transport is purely determined by the scattering matrix, all models that produce the same scattering 
matrix give identical results.
Regardless of actual microscopic detail, we may therefore conveniently take the Hamiltonian of the scatterer to be
\begin{align}
\mathcal H=&v_F\sum_{m,n}\int dz\; \psi^\dagger_m(z)\left\{-i\delta_{m,n}\partial_z+u_{m,n}(z)\right\}\psi_n(z)\nonumber\\
&+ \mathcal H_{\rm res} + \mathcal H_{\rm T},\label{genham}
\end{align}
where $\mathcal H_{\rm res}$ represents the reservoirs, and $\mathcal H_{\rm T}$ takes account of tunneling between
the conductor and the reservoirs. The scattering region and the reservoirs are spatially separated. This means
that the scattering potential $u_{mn}(z)$ is non-zero only in a region $z^-<z<z^+$ while tunneling between
the reservoirs and the conductor only takes place outside this region.
Note that in our model, scattering channels have been ``unfolded'',
so that in stead of working with a channel that confines particles in the interval $(-\infty,0]$ and allowing for
propagation both in the positive and negative directions, we equivalently work with channels in which particles propagate along
$(-\infty,\infty)$, but only in the positive direction. Hence, to make contact with most physical setups, we consider
$-z$ and $z$ to refer to the same physical position in a channel, but opposite propagation directions.

We consider the generating functional 
\begin{widetext}
\begin{equation}
\mathcal Z=e^{\mathcal A}={\rm Tr}\left[\mathcal T^+\exp\left\{-i\int_{t_0}^{t_1}dt\;\mathcal H^+(t)\right\}
\rho_0\mathcal T^-\exp\left\{i\int_{t_0}^{t_1}dt\;\mathcal H^{-}(t)\right\}\right]
\end{equation}
\end{widetext}
in which $\mathcal H^\pm$ is obtained from $\mathcal H$ by replacing $u_{mn}(z)$ with arbitrary time-dependent
functions $u^\pm_{mn}(z,t)$. 
In this expressions $\mathcal T^+\exp$ and $\mathcal T^-\exp$ respectively refer to time-ordered (i.e. largest time to the left)
and anti-time-ordered (i.e. largest time to the right) exponentials. 
In the language of Feynman end Vernon \cite{Fey63} this is known as the influence functional. It gives a complete
characterization of the effect that the electrons in the scatterer have on any quantum system that interact with. 
Furthermore, the functional
$\mathcal Z$ generates expectation values of time-ordered products of operators as follows. 
Let $Q$ be an operator
\begin{equation}
Q=\sum_{mn}\int_{z^-}^{z^+}dz\,\psi_m^\dagger(z) q_{mn}(z)\psi_n(z).
\end{equation}
Choose $u^\pm_{mn}(z,t)=u_{mn}(z)+\chi_{\pm}(t)q_{mn}(z)$. Then
\begin{align}
&\left<\mathcal T^-\left(\prod_{j=1}^M Q(t_j)\right)
\mathcal T^+\left(\prod_{k=1}^N Q(t_k')\right)\right>\nonumber\\
&=\prod_{j=1}^M\left(-i\frac{\delta}{\delta\chi^-(t_j)}\right)
\prod_{k=1}^N\left(i\frac{\delta}{\delta\chi^+(t_k')}\right)\left.\mathcal Z[\chi]\right|_{\chi=0}
\end{align}
By merging the power of the Keldysh formalism of contour-ordered Green functions with that of the Landauer scattering
formalism for quantum transport, we obtain an expression for $\mathcal Z$ in terms of the Keldysh Green functions
in the reservoirs and the time dependent scattering matrices associated with $\hat{u}^\pm(z,t)$.

The argument will proceed in the following steps:
\begin{enumerate}
\item{Firstly we introduce the key object that enables a systematic analysis of $\mathcal Z$, namely the single particle 
Green function $g$ of the conductor. We state the equations of motion that $g$ obeys.}
\item{We define the Keldysh action $\mathcal A=\ln\mathcal Z$, and consider its variation $\delta \mathcal A$. 
We discover that $\delta \mathcal A$ can be expressed in terms of $g$.}
\item{We therefore determine $g$ inside the scattering region in terms of the scattering matrix of the conductor and
its value at the edges of the scattering region, where the reservoirs impose boundary conditions.}
\item{This allows us to express the variation of the action in terms of the reservoir Green functions $G_{\rm in\,(out)}$
and the scattering matrix $s$ of the conductor.}
\item{The variation $\delta \mathcal A$ is then integrated to find the action $\mathcal A$ and the 
generating functional $\mathcal Z$.}
\end{enumerate} 

\subsection{Preliminaries: Definition of the Green function}
The first step is to move from the Schr\"odinger picture to the Heisenberg picture. 
To shorten notation we define two time-evolution operators:
\begin{equation}
\mathcal U_\pm(t_f,t_i)=\mathcal T^+\exp\left\{-i\int_{t_i}^{t_f}dt'\;\mathcal H^\pm(t')\right\}.
\end{equation}
Associated with every Schr\"odinger picture operator we define two Heisenberg operators, one
corresponding to evolution with each of the two Hamiltonians $\mathcal H^\pm$.
\begin{equation}
Q_\pm(t)=\mathcal U_\pm(t_f,t_i)^\dagger \,Q\,\mathcal U_\pm(t_f,t_i).
\end{equation}
\begin{widetext}
In order to have the tools of the Keldysh formalism at our disposal, we need to define four Green functions
\begin{eqnarray}
g_{m,n}^{++}(z,t;z',t')&=&-e^{\mathcal A}{\rm Tr}\left[\mathcal U^+(t_1,t_0)\mathcal T^+\left(\psi_{n+}^\dagger(z',t')\psi_{m+}(z,t)\right)
\rho_0\left(\mathcal U^-(t_1,t_0)\right)^\dagger\right]\nonumber\\
g_{m,n}^{+-}(z,t;z',t')&=&e^{\mathcal A}{\rm Tr}\left[\mathcal U^+(t_1,t_0) \psi_{m+}(z,t)
\rho_0\psi^\dagger_{n-}(z',t')\left(\mathcal U^-(t_1,t_0)\right)^\dagger\right]\nonumber\\
g_{m,n}^{-+}(z,t;z',t')&=&e^{\mathcal A}{\rm Tr}\left[\mathcal U^+(t_1,t_0) \psi_{n+}^\dagger(z',t')
\rho_0\psi_{m-}(z,t)\left(\mathcal U^-(t_1,t_0)\right)^\dagger\right]\nonumber\\
g_{m,n}^{--}(z,t;z',t')&=&e^{\mathcal A}{\rm Tr}\left[\mathcal U^+(t_1,t_0)
\rho_0\mathcal T^-\left(\psi_{n-}^\dagger(z',t')\psi_{m-}(z,t)\right)\left(\mathcal U^-(t_1,t_0)\right)^\dagger\right].
\end{eqnarray}
\end{widetext}
Here the symbol $\mathcal T^+$ orders operators with larger time arguments to the left. 
If permutation is required to obtain the time-ordered form, the product is
multiplied with $(-1)^n$ where $n$ is the parity of the permutation. 
Similarly, $\mathcal T^-$ anti-time-orders with the same permutation parity convention.

The Green functions can be grouped into a matrix in Keldysh space
\begin{equation}
g_{m,n}(z,t;z',t')=\left(\begin{array}{cc}g^{++}_{m,n}(z,t;z',t')&g^{+-}_{m,n}(z,t;z',t')
\\g^{-+}_{m,n}(z,t;z',t')&g^{--}_{m,n}(z,t;z',t')\end{array}\right)\label{eqgf}.
\end{equation}
Notation can be further shortened by incorporating channel-indices 
into the matrix structure of the Green function, thereby defining an object $\bar g(z,t;z',t')$. The
element of $\bar g$ that is located on row $m$ and column $n$, is the $2\times2$ matrix $g_{m,n}$.

The Green function satisfies the equation of motion
\begin{align}
&\left\{i\partial_t+v_F i\partial_z-v_F\bar u(z,t)\right\}\bar g(z,t;z',t')\nonumber\\
&-\int dt''\Sigma(z;t-t'')\bar g(z,t'';z't')=\delta(t-t')\delta(z-z')\bar 1.\label{eqmot}
\end{align}
The delta-functions on the right of Eq. (\ref{eqmot}) encode the fact that due to 
time-ordering $g_{mn}^{++}$ and $g_{mn}^{--}$ have a step-structure
\begin{equation}
\frac{1}{v_F}\theta(z-z')\delta(t-t'-\frac{z-z'}{v_F})\delta_{mn}+f(z,t;z't')
\end{equation}
where $f$ is continuous in all its arguments.
The self-energy 
\begin{equation}
\Sigma(z;\tau)=-i\frac{\bar G_{\rm in}(\tau)}{2\tau_c}\theta(z^--z)
-i\frac{\bar G_{\rm out}(\tau)}{2\tau_c}\theta(z-z^+)
\end{equation} 
results from the reservoirs and determines how
the scattering channels are filled. It is a matrix in Keldysh space. 
The time $\tau_c$ is the characteristic time correlations survive in the region of the
conductor that is connected to the reservoirs, before the reservoirs scramble them. 
$\bar G_{\rm in\,(out)}(\tau)$ is the reservoir Green functions where electrons enter (leave) the scattering region, 
summed over reservoir levels and normalized to be dimensionless. 
This form of the self-energy can be derived from the following model for the reservoirs: 
We imagine every point $z$ in a channel $m$ outside $(z^-,z^+)$ to
exchange electrons with an independent Fermion bath with a constant density of 
states $\nu$. The terms $\mathcal H_{\rm res}$ and $\mathcal H_{\rm T}$ are explicitly
\begin{eqnarray}
\mathcal H_{\rm res}&=&\sum_m\int dE\,\nu\,\int_{z\not\in(z^-,z^+)}dz\,E\, a_m^\dagger(E,z)a_m(E,z)\nonumber\\    
\mathcal H_{\rm T}&=&\sum_m\,c_m\int dE\,\nu\,\int_{z\not\in(z^-,z^+)}dz\,
\psi^\dagger_m(z)a_m(E,z)\nonumber\\
&&\hspace{5mm}+ a^\dagger_m(E,z)\psi_m(z), 
\end{eqnarray}
where the tunneling amplitude $c_m$ characterizes the coupling between the reservoir and channel $m$.
More general reservoir models need not be considered, since, as we shall see shortly, the 
effect of the reservoirs is contained entirely in a boundary conditions on the Green function $\bar g$ 
inside the scatterer. This boundary condition does not depend on microscopic detail, but
only on the reservoir Green functions $\bar G_{\rm in\,(out)}$.  

We do not need to know the explicit form of the 
reservoir Green functions yet. Rather the argument below relies exclusively on the
property of $\bar G_{\rm in\,(out)}$ that it squares to unity \cite{Ram86}:
\begin{equation}
\int dt''\,\bar G(t-t'')_{\rm in\,(out)}\bar G(t''-t')_{\rm in\,(out)}=\delta(t-t')\bar 1.
\end{equation}
A differential equation similar to Eq. (\ref{eqmot}) holds for $\bar{g}^\dagger$. 

\subsection{Varying the action $\mathcal A$.}
We are now ready to attack the generating functional $\mathcal Z$. For our purposes, it is most convenient 
to consider $\mathcal A=\rm ln\,\mathcal Z$. We will call this object the action. Our strategy is as follows:
We will obtain an expression for the variation $\delta \mathcal A$ resulting from a variation 
$\hat{u}(z,t)\to \hat{u}(z,t)+\delta \hat{u}(z,t)$ of the scattering potentials. 
This expression will be in terms of the reservoir filling factors $\hat{f}$ and the scattering matrices 
associated with $\hat{u}(z,t)$. We then integrate to find $\mathcal A$.

\begin{widetext}
We start by writing 
\begin{equation}
\delta \mathcal A=-i v_F e^{\mathcal A}\sum_{m,n}\int_{t_0}^{t_1} dt\int dz\;\left(\delta u^+_{n,m}(z,t)\left<\psi_m^\dagger(z)\psi_n(z)\right>_+(t)-
\delta u^-_{n,m}(z,t)\left<\psi_m^\dagger(z)\psi_n(z)\right>_-(t)\right)\label{vara}
\end{equation}
where
\begin{eqnarray}
\lefteqn{\left<\psi_m^\dagger(z)\psi_n(z)\right>_+(t)}\nonumber\\
&=&{\rm Tr}\left[\mathcal T^+\exp\left\{-i\int_{t}^{t_1}dt'\;\mathcal H^+(t')\right\}
\psi_m^\dagger(z)\psi_n(z)\mathcal T^+\exp\left\{-i\int_{t_0}^tdt'\;\mathcal H^+(t')\right\}
\rho_0\mathcal T^-\exp\left\{i\int_{t_0}^{t_1}dt'\;\mathcal H^{-}(t')\right\}\right]\nonumber\\
\lefteqn{\left<\psi_m^\dagger(z)\psi_n(z)\right>_-(t)}\nonumber\\
&=&{\rm Tr}\left[\mathcal T^+\exp\left\{-i\int_{t_0}^{t_1}dt'\;\mathcal H^+(t')\right\}
\rho_0\mathcal T^-\exp\left\{i\int_{t_0}^tdt'\;\mathcal H^{-}(t')\right\}
\psi_m^\dagger(z)\psi_n(z)\mathcal T^-\exp\left\{-i\int_{t}^{t_1}dt'\;\mathcal H^-(t')\right\}\right].\nonumber\\
\end{eqnarray}
\end{widetext}
\subsection{Expressing $\delta \mathcal A$ in terms of the Green function $g$.}
In terms of the defined Green functions, the variation $\delta \mathcal A$ becomes
\begin{eqnarray}
\delta \mathcal A&=&i v_F \sum_{m,n}\int_{t_0}^{t_1} dt\int dz\ \; \Big(\delta u^+_{n,m}(z,t)\nonumber\\
&&\hspace{15mm}\times g_{m,n}^{++}(z,t-0^+;z,t)\nonumber\\
&&\hspace{1cm}+\delta u^-_{n,m}(z,t)g_{m,n}^{--}(z,t+0^+;z,t)\Big)\nonumber\\
&=&i v_F \int_{t_0}^{t_1} dt\int dz\;{\rm Tr}\left[\delta \bar u(z,t)\bar g(z,t+0^k;z,t)\right].\nonumber\\
\label{trvar}
\end{eqnarray}
The object $\delta \bar u$ is constructed by combining the channel and Keldysh indices of the variation of the potential. 
The trace is over both Keldysh and
channel indices. The symbol $0^k$ refers to the regularization explicitly indicated in the first line, i.e. 
the first time argument of $g^{++}(z,t-0^+;z,t)$ is evaluated
an infinitesimal time $0^+>0$ {\em before} the second argument, while in $g^{--}(z,t+0^-;z,t)$, 
the first time argument is evaluated an infinitesimal time $0^+$ {\em after}
the second. This is done so that the time ordering (anti-time ordering) operations give 
the order of creation and annihilation operators required in Eq. (\ref{vara}).

It proves very inconvenient to deal with the $0^k$ regularization of Eq. (\ref{trvar}). It is preferable to have the first time arguments of
both $g^{++}$ and $g^{--}$ evaluated an infinitesimal time $0^+$ {\em before} the second. Taking into account the step-structure 
of $\hat g^{++}$ we have
\begin{align}
&\bar g(z,t+0^k;z't')=\bar g(z,t-0^+;z',t')\nonumber\\
&\hspace{5mm}+\frac{1}{v_F}\delta(t-t'-\frac{z-z'}{v_F})\hat 1\left(\frac{1-\check \tau_3}{2}\right).
\end{align}
Here $\check \tau_3$ is the third Pauli matrix $\left(\begin{array}{cc}1&0\\0&-1\end{array}\right)$ acting in Keldysh space.
The equations of motion allow us to relate $\bar{g}(z,t-0^+;z',t')$ for points $z$ and $z'$ inside the scattering 
region where $\bar u$ is non-zero, to the value of $\bar g$ at $z^-$ where electrons enter the scatterer.
For $z\leq z'$ and $t\leq t'$, the equations of motion give
\begin{align}
&\bar{g}(z,t+\frac{z-z^-}{v_F}-0^;z',t+\frac{z'-z^-}{v_F})\nonumber\\
&=\bar s(z,t) \bar g(z^-,t-0^+;z'^-,t')\bar s^\dagger(z',t'),\label{sol}
\end{align}
where
\begin{equation}
\bar s(z,t)=\mathcal Z \exp\left\{-i\int_{z^-}^{z}dz''\bar u(z'',t+\frac{z''-z^-}{v_F})\right\}.
\end{equation} 
The symbol $\mathcal Z$ indicates that the exponent is ordered along the $z$-axis, with the largest co-ordinate in the integrand 
to the left. Note that the potential $\bar u$ at position $z$ is evaluated at the time instant $t+(z-z^-)/v_F$
that an electron entering the scattering region at time $t$ reaches $z$.
Often the time-dependence of the potential is slow on the time-scale $(z^+-z^-)/v_F$ representing the time a 
transported electron spends in the scattering region and $\bar u(z,t+\frac{z-z^-}{v_F})$ can be replaced 
with $\bar u(z,t)$. This is however not required for the analysis that follows to be valid.
 
Substitution into Eq. (\ref{trvar2}) yields
\begin{align}
\delta A=&v_F\int dt\;{\rm Tr}\left[\bar w(t) g(z^-,t-0^+;z^-,t)\right]\nonumber\\
&-\int dt\; \lim_{t'\rightarrow t} \delta(t-t')
{\rm Tr}\;\left[\bar w(t)\hat 1\left(\frac{1-\check \tau_3}{2}\right)\right].\label{trvar2}
\end{align}
with 
\begin{eqnarray}
\bar{w}(t)&=&-i\int_{z^-}^{z^+}dz \bar s^\dagger(z,t) \delta \bar u(z,t+\frac{z-z^-}{v_F}) \bar s(z,t)\nonumber\\
&=&\bar s^\dagger(t) \delta \bar s(t).\label{eqw}
\end{eqnarray}
In this equation $z^+$ is located where electrons leave the scatterer. 
Importantly, here $\rm Tr$ still denotes a trace over channel and Keldysh indices. We will later on redefine the symbol to include also a trace
over the (continuous) time index, at which point the second term in Eq. (\ref{trvar2}) will (perhaps deceptively) look less offensive, but not yet.
In the last line of Eq. (\ref{eqw}), $\bar s(t)=\bar s(z^+,t)$ is the (time-dependent) scattering matrix. 
We sent the boundaries $t_0$ and $t_1$ over which we integrate in the definition of the action, to $-\infty$
and $\infty$ respectively, which will allow us to Fourier transform to frequency in a moment. 
The action remains well-defined as long as the potentials $u^+$ and $u^-$ only differ for a finite time.

\subsection{Relating $g$ inside the scattering region to $g$ at reservoirs. Imposing boundary conditions implied by reservoirs.}
Our task is now to find $\bar g(z^-,t-0^+;z^-,t)$. 
Because of the $t-t'$ dependence of the self-energy, it is convenient to transform to Fourier space, where
\begin{eqnarray}
\bar g(z,\varepsilon;z^-,\varepsilon')&=&\int dt\,dt'\,e^{i\varepsilon t}\bar g(z,t;z^-,t')
e^{-i\varepsilon' t'}\nonumber\\
\bar G_{\rm in\,(out)}(\varepsilon)&=&\int dt\; e^{i\varepsilon t} \bar G(t)_{\rm in\,(out)}.
\end{eqnarray}
In frequency domain, the property that $\bar G_{\rm in\,(out)}$ squares to unity is expressed as
$\bar G_{\rm in\, (out)}(\varepsilon)^2=\bar 1$. (Due to the standard conventions for Fourier transforms,
the matrix elements of the identity operator in energy domain is $2\pi \delta(\varepsilon-\varepsilon')$.)
The equation of motion for $z<z^-$ reads
\begin{equation}
\left\{-i\varepsilon+v_F\partial_z+\frac{\bar G_{\rm in}(\varepsilon)}{2\tau_c}\right\}\bar g(z,\varepsilon;z^-,\varepsilon')=0.
\end{equation}
There is no inhomogeneous term on the right-hand side, because we restrict $z$ to be less than $z^-$. We thus find
\begin{align}
&\bar g(z^--0^+,\varepsilon;z^-, \varepsilon')\nonumber\\
&=e^{i\varepsilon\Delta z/v_F}\exp\left[-\frac{\bar G_{\rm in}(\varepsilon)}{2l_c}
\Delta z\right]\bar g(z^--\Delta z,\varepsilon;z^-,\varepsilon').
\end{align}
Here the correlation length $l_c$ is the correlation time $\tau_c$ multiplied by the Fermi velocity $v_F$.
Using the fact that $\bar G(\varepsilon)_{\rm in}$ squares to unity, it is easy to verify that
\begin{align}
&\exp\left\{-\frac{\bar G_{\rm in}(\varepsilon)}{2l_c}\Delta z\right\}\nonumber\\
&=\frac{1+\bar G_{\rm in}(\varepsilon)}{2}
\exp\left(-\frac{\Delta z}{2 l_c}\right)+
\frac{1-\bar G_{\rm in}(\varepsilon)}{2}\exp\left(\frac{\Delta z}{2 l_c}\right).
\end{align}
Since spacial correlations decay beyond $z^-$, $\bar g(z^--\Delta z,\varepsilon;z^-,\varepsilon')$
does not blow up as we make $\Delta z$ larger.
From this we derive the condition
\begin{equation}
\left[1+\bar G_{\rm in}(\varepsilon)\right]\bar g(z^--0^+,\varepsilon;z^-,\varepsilon')=0.
\end{equation}
\begin{widetext}
Transformed back to the time-domain this reads
\begin{equation}
\int dt''\;\left[\delta(t-t'')+\bar G_{\rm in}(t-t'')\right]\bar g(z^--0^+,t'';z^-,t')=0.\label{eqgm}
\end{equation}

We can play the same game at $z^+$ where particles leave the scatterer.  
The equation of motion reads
\begin{equation}
\left\{-i\varepsilon+v_F\partial_z+\theta(z-z^+)\frac{\bar G_{\rm out}(\varepsilon)}{2\tau_c}\right\}\bar g(z,\varepsilon;z^+,\varepsilon')
=2\pi\delta(z-z')\delta(\varepsilon-\varepsilon').
\end{equation}
This has the general solution
\begin{eqnarray}
\bar g(z,\varepsilon;z',\varepsilon')&=&
\exp\left\{i\varepsilon\frac{z-z'}{v_F}-\left[(z-z^+)\theta(z-z^+)-(z'-z^+)\theta(z'-z^+)\right]\frac{\bar G_{\rm out}(\varepsilon)}{2l_c}
\right\}\nonumber\\
&&\hspace{1cm}\times\left[\bar g(z'-0^+,\varepsilon';z',\varepsilon')+\frac{2\pi}{v_F}\theta(z-z')\delta(\varepsilon-\varepsilon')\right].
\end{eqnarray}
We will need to relate the Green function evaluated at $z<z^+$ to the Green function evaluated at $z>z^+$, and so we explicitly show the inhomogeneous
term. The same kind of argument employed at $z^-$ then yields the condition
\begin{equation}
\left[1-\bar G_{\rm out}(\varepsilon)\right]\left[\bar g(z^+-0^+,\varepsilon;z^+,\varepsilon')+\frac{2\pi}{v_F}\delta(\varepsilon
-\varepsilon')\right]=0,
\label{rb}
\end{equation}
where the inhomogeneous term in the equation of motion is responsible for the delta-function. In time-domain this reads
\begin{equation}
\int dt''\;\left[\delta(t-t'')-\bar G_{\rm out}(t-t'')\right]\left[\bar g(z^+-0^+,t'';z^+,t')+\frac{1}{v_F}\delta(t''-t')\right]=0.\label{eqgp}
\end{equation}

It remains for us to relate $\bar g(z^+-0^+,t+\frac{z^+-z^-}{v_F};z^+,t'+\frac{z^+-z^-}{v_F})$ to $\bar g(z^--0^+,t;z^-,t')$. 
This is done with the help of Eq. (\ref{sol}), from which follows
\begin{equation}
\bar g(z^+-0^+,t+\frac{z^+-z^-}{v_F};z^+,t'+\frac{z^+-z^-}{v_F})=\bar s(t)\bar g(z^--0^+,t;z^-,t') \bar s^\dagger(t').
\label{eq1}
\end{equation}
We substitute this into Eq. (\ref{eqgp}), multiply from the right with $\bar s(t')$ and from the left with $\bar s^\dagger(t)$. If we define
$\bar G'_{\rm out}(t,t')=\bar s^\dagger(t) \bar G_{\rm out}(t-t')\bar s(t')$ the resulting boundary condition is
\begin{equation}
\int dt''\;\left[\delta(t-t'')-\bar G'_{\rm out}(t-t'')\right]\left[\bar g(z^--0^+,t'';z^-,t')+\frac{1}{v_F}\delta(t''-t')\right]=0.\label{eqgp1}
\end{equation}
\end{widetext}
\subsection{Finding the variation of the action in terms of the reservoir Green functions and the scattering matrix.}
At this point, it is convenient to incorporate time into the matrix-structure of the objects $\bar G_{\rm in}$, $\bar G'_{\rm out}$
and $\bar g$. The resulting matrices will be written without overbars. Thus for instance $s$ will denote a matrix diagonal in time-indices, whose
entry $(t,t')$ is $\delta(t-t')\bar s(t)$. Similarly the $(t,t')$ entry of $G_{\rm in\,(out)}$ is $\bar G_{\rm in\,(out)}(t-t')$. 
Also let $g^-$ be the matrix whose $(t,t')$
entry is $\bar g(z^--0^+,t;z^-,t')$. In this notation $G_{\rm in}^2=\left.G'_{\rm out}\right.^2=I$ and Eq. (\ref{eqgm}) 
and Eq. (\ref{eqgp1}) read
\begin{eqnarray}
\left(I+G_{\rm in}\right)g^-&=&0\nonumber\\
\left(I-G'_{\rm out}\right)\left(g^-+1/v_F\right)&=&0.\label{boun}
\end{eqnarray}
These two equations determine $g^-$ uniquely as follows: From the first of the two equations we have
\begin{eqnarray}
0&=&G'_{\rm out}(I+G_{\rm in})g^-\nonumber\\
&=&-(I-G'_{\rm out})g^-+(I+G'_{\rm out}G_{\rm in})g^-
\end{eqnarray}
In the first term we can make the substitution $-(I-G'_{\rm out})g^-=(I-G'_{\rm out})/v_F$ which follows from Eq. (\ref{boun}). Thus we find
\begin{eqnarray}
g^-&=&-\frac{1}{v_F}\frac{1}{I+G'_{\rm out}G_{\rm in}}(I-G'_{\rm out})\nonumber\\
&=&\frac{1}{v_F}(1-G_{\rm in})\frac{1}{G'_{\rm out}+G_{\rm in}}
\end{eqnarray}
and the last line follows from the fact that $G_{\rm in}^2=\left.G'_{\rm out}\right.^2=I$.
We have taken special care here to allow for different reservoir Green functions at $z^-$ where
particles enter the conductor and $z^+$ where they leave the conductor. In order to proceed we
must now absorb the difference between the two Green functions in the scattering matrix. 
We define $\Lambda$ through the equation
\begin{equation}
\bar G_{\rm out}=\Lambda^{-1} G_{\rm in} \Lambda
\end{equation}
and drop subscripts on the Green functions by setting $G\equiv G_{\rm in}$.
Substituted back into Eq. (\ref{trvar2}) for the variation of the action yields
\begin{equation}
\delta \mathcal A={\rm Tr}\left[\delta s'(1-G)\frac{1}{Gs'+s'G}\right]-{\rm Tr}\left[\delta \hat s_-(\hat s_-)^\dagger\right],
\end{equation}
where the trace is over time, channel and, in the first term, Keldysh indices. The operator $s'$ is related to the scattering
matrix $s$ through $s'=\Lambda s$.
\subsection{Integrating the variation to find the action $\mathcal A$.}
We now have to integrate $\delta \mathcal A$ to find $\mathcal A$. This is most conveniently done by working in a basis where $G$ is
diagonal. Since $G^2=1$, every eigenvalue of $G$ is $\pm 1$. Therefore, there is a basis in which
\begin{equation}
G=\left(\begin{array}{cc}I&0\\0&-I\end{array}\right).
\end{equation}
In this representation $s'$ can be written as 
\begin{equation}
s'=\left(\begin{array}{cc}s_{11}'&s_{12}'\\s_{21}'&s_{22}'\end{array}\right).
\end{equation}
Here the two indices of the subscript has the following meaning: The first refers to a left eigenspace of $G$,
the second to a right eigenspace. A subscript $1$ denotes the subspace of eigenstates of $G$ with eigenvalue $1$.
A subscript $2$ refers to the subspace of eigenstates of $G$ with eigenvalue $-1$. In this representation,
\begin{equation}
(1-G)\frac{1}{Gs'+s'G}=\left(\begin{array}{cc}0&0\\0&(s'_{22})^{-1}\end{array}\right),
\end{equation}
so that 
\begin{equation}
\delta \mathcal A={\rm Tr}\left[\delta s_{22}'\; (s_{22}^{-1})'\right]-{\rm Tr}\left[\delta \hat s_-(\hat s_-)^\dagger\right],
\end{equation}
and thus
\begin{eqnarray}
\mathcal A&=&{\rm Tr\;ln}\;s_{22}'-{\rm Tr\;ln}\; s_-\nonumber\\
e^{\mathcal A}&=&\left({\rm Det}\; s_-\right)^{-1}{\rm Det}\;s_{22}'\label{consts}
\end{eqnarray}
In these equations, $s_-$ is the scattering matrix associated with $\mathcal H^-$ as defined previously. Its time structure
is to be included in the operations of taking the trace and determinant. 

Note that in the representation where $G$ is diagonal, it holds that
\begin{equation}
\frac{1+G}{2}+s'\frac{1-G}{2}=\left(\begin{array}{cc}I&s_{12}'\\0&s_{22}'\end{array}\right).
\end{equation}
Due to the upper-(block)-triangular structure it holds that ${\rm Det}\;s_{22}'={\rm Det}\left[\frac{1+G}{2}+s'\frac{1-G}{2}\right]$ 
leading to our main result
\begin{equation}
\mathcal A={\rm Tr\;ln}\left[\frac{1+G}{2}+s'\frac{1-G}{2}\right]-{\rm Tr\;ln}\;s_-.\label{firstform}
\end{equation}
where it has to be noted that many matrices have the same determinant as the above. Some obvious examples include
\begin{eqnarray}
\left(\begin{array}{cc}I&0\\0&s_{22}'\end{array}\right)&=&(1+G)/2+(1-G)s'(1-G)/4\nonumber\\
\left(\begin{array}{cc}I&0\\s_{21}'&s_{22}'\end{array}\right)&=&(1+G)/2+(1-G)s'/2.
\end{eqnarray}
\section{Tracing out the Keldysh structure}
Up to this point the only property of $G$ that we relied on was the fact that it squares to identity. 
Hence the result (Eq. \ref{firstform}) holds in a setting that is more general than that of
a scatterer connected to reservoirs characterized by filling factors. (The reservoirs may 
for instance be superconducting). In the specific case of reservoirs characterized by filling factors it
holds that
\begin{equation}
\bar G(\tau)=\int \frac{d\varepsilon}{2\pi} e^{-i\varepsilon\tau}\left(\begin{array}{ll}1-2\hat f(\varepsilon)&2\hat f(\varepsilon)\\
2-2\hat f(\varepsilon)&-1+2\hat f(\varepsilon)\end{array}\right).\label{defG}
\end{equation}
Here $\hat f(\epsilon)$ is diagonal in channel indices, and $f_m(\epsilon)$ is the filling factor in the reservoir
connected to channel $m$. We will also assume that electrons enter and leave a channel from the same reservoir, 
so that $G_{\rm in}=G_{\rm out}$ and hence $s'=s$. We recall as well as that the Keldysh structure of  
the scattering matrix is
\begin{equation}
s=\left(\begin{array}{cc}\hat s_+&0\\0&\hat s_-\end{array}\right).
\end{equation}
Here $\hat s_\pm$ have channel and time (or equivalently energy) indices. $\hat s_\pm$ is diagonal in time-indices, with the entries on the time-
diagonal the time-dependent scattering matrices corresponding to evolution with the Hamiltonians $\mathcal H_\pm$. 

With this structure in Keldysh space, we find
\begin{align}
e^{\mathcal A}=&{\rm Det}\left(\begin{array}{ll} 1+(\hat s_+ -1)\hat f&-(\hat s_+-1)\hat f\\
(\hat s_--1)(\hat f-1)&\hat s_-(1-\hat f)+\hat f\end{array}\right)\nonumber\\
&\times{\rm Det}\left(\begin{array}{ll}1&\\&\hat{s}_-^{-1}\end{array}\right).
\end{align}
We can remove the Keldysh structure from the determinant with the aid of the general formula
\begin{eqnarray}
{\rm Det}\left(\begin{array}{cc}A&B\\C&D\end{array}\right)&=&{\rm Det}(AD-ACA^{-1}B)\nonumber\\
&=&{\rm Det}(DA-CA^{-1}BA).
\end{eqnarray}
Noting that in our case the matrices $B$ and $A$ commute, so that $CA^{-1}BA=CB$, we have
\begin{widetext}
\begin{eqnarray}
e^{\mathcal A}&=&{\rm Det}\left[\left(\hat s_-(1-\hat f)+\hat f\right)\left(1+(\hat s-1)\hat f\right)-\left(\hat s_-(1-\hat f)+\hat f-1\right)
(\hat s_+-1)\hat f\right]\,{\rm Det}\left(\hat{s}_-^{-1}\right)
\nonumber\\
&=&{\rm Det}\left[\hat s_-(1-\hat f)+\hat s_+\hat f\right]{\rm Det}\left(\hat{s}_-^{-1}\right).\label{mainres}
\end{eqnarray}
\end{widetext}

\section{An example: Full Counting Statistics of transported charge.}
A determinant formula of this type appears in the literature of Full Counting Statistics \cite{Kli03} of transported charge.
This formula can be stated as follows: the generating function for transported charge through 
a conductor characterized by a scattering matrix $\hat s$ is
\begin{equation}
\mathcal Z[\chi]={\rm Det}\left[1+(\hat s^\dagger_{-\chi} \hat s_\chi - 1) \hat f\right]
\end{equation}
where $\hat s_\chi$ is a scattering matrix, modified to depend on the counting field $\chi$
that, in this case, is time-independent. (The
precise definition may be found below.)  

As a consistency check of our results, we apply our analysis to re-derive
this formula. We will consider the most general setup, where every scattering channel is connected
to a distinct voltage-biased terminal. To address the situation where leads connect several channels to the same terminal,
the voltages and ``counting fields'' associated with channels in the same lead, are set equal.

Let us start by defining number operators
\begin{equation}
\mathcal N_k=\int dz\;\left[\theta(-z_0-z)+\theta(z-z_0)\right]\psi_k^\dagger(z)\psi_k(z)
\end{equation}
The operator $\mathcal N_k$ counts the number of particles in channel $k$ that are located outside the interval $[-z_0,z_0]$.
The coordinate $z_0$ is chosen to lie between the reservoirs the scattering region. 
With the full counting statistics of transported charge, we mean the generating functional (a la Levitov)
\begin{align}
&\mathcal Z(\chi_1,\ldots,\chi_N,t)\nonumber\\
&=\left<e^{i\sum_k\chi_k\mathcal N_k(0)/2}
e^{-i\sum_k\chi_k\mathcal N_k(t)}e^{i\sum_k\chi_k\mathcal N_k(0)/2}\right>
\end{align}
The function $\mathcal Z(\chi,t)$ generates all moments of the joint distribution function for $(n_1,n_2,\ldots,n_N)$ charges to be 
transported into terminal $(1,2,\ldots,N)$ in the time interval $(0,t)$. Formally $t$ is sent to infinity, and this causes
a singularity in all irreducable moments of the distribution of thransported charge.
Dealing with this singularity is a subtle issue, which 
we will not concern ourselves with. The interested reader is referred to the literature \cite{Iva97,Avr07}.

The dynamics of the the operators $\mathcal N_k(t)$ are determined by the Hamiltonian
\begin{equation}
\mathcal H=v_f\sum_{m,n}\int dz\;\psi_m^\dagger(z)\left\{-i\partial_z\delta_{m,n}+u_{m,n}(z)\right\}\psi_n(z).
\end{equation}
As already mentioned, we choose to count charges outside the scattering potential region, so that every
$\mathcal N_k$ commutes with the potential energy term in the Hamiltonian. Using the short-hand notation
$\sum_m \chi_m\mathcal N_m=\chi.\mathcal N$, we manipulate the definition of $\mathcal Z(\chi,t)$ to find
\begin{eqnarray}
\mathcal Z(\chi,t)&=&\left<e^{i\chi.\mathcal N/2}e^{i\mathcal H t}e^{-i\chi.\mathcal N}
e^{-i\mathcal H t}e^{-i\chi.\mathcal N/2}\right>\nonumber\\
&=&\left<e^{i\mathcal H_\chi t}e^{-i\mathcal H_{-\chi} t}\right>.
\end{eqnarray}
In this equation, the Hamiltonian $\mathcal H_\chi$ is defined as
\begin{align}
&\mathcal H_\chi=e^{i\chi.\mathcal N/2}\mathcal H e^{-i\chi.\mathcal N/2}\nonumber\\
&=v_f\sum_{m,n}\int dz\;\psi_m^\dagger(z)\left\{-i\partial_z\delta_{m,n}+u^{(\chi)}_{m,n}(z)\right\}\psi_n(z)\nonumber\\
\end{align}
The transformed potential is $u_{m,n}^{(\chi)}(z)=u_{m,n}(z)+\delta_{m,n}\frac{\chi_m}{2}(\delta(z-z_0)-\delta(z+z_0))$. One way to
verify this is to note that formally $\mathcal H_\chi$ is related to $\mathcal H$ through a gauge transformation where
the gauge field in each channel is proportional to
$\theta(z-z_0)+\theta(-z-z_0)$. The delta-functions in $u^{(\chi)}$ arise as a gradient of the gauge field that appear
in the transformation of the kinetic term in $\mathcal H$.

The calculation of the full counting statistics has now been cast into the form of the trace of a density matrix after forward and
backward time evolution controlled by different scattering potentials. Our result, Eq. (\ref{mainres}), is therefore applicable, with
\begin{eqnarray}
\hat s_\pm&=&\mathcal Z{\rm exp}\left(-i\int_{z_-}^{z_+} dz\;\hat u^{(\pm\chi)}(z)\right)\nonumber\\
&=&e^{\mp i\hat\chi/2}s_0e^{\pm i\hat\chi/2}=s_{\pm\chi}.
\end{eqnarray}
In this equation, $\hat \chi$ is a diagonal matrix in channel space, with entries $\delta_{m,n}\chi_m$. 
Substitution into
Eq. (\ref{mainres}) gives
\begin{equation}
\mathcal Z(\chi)={\rm Det}\left[1+(\hat s_{-\chi}^\dagger \hat s_\chi-1)\hat f\right],\label{fcs}
\end{equation}
in agreement with the existing literature\cite{Kli03}. 

\section{Tracing out the channel structure.}
A large class of experiments and devices in the field of quantum transport is based on two terminal setups.
In such a setup the channel space of the scatterer is naturally partitioned into a left and right set,
each connected to its own reservoir. 
We are generally interested in transport between left and right as opposed to internal 
dynamics on the left- or right-hand sides.
The scattering matrices have the general structure
\begin{equation}
\hat s_\pm=X
\left(\begin{array}{cc} r& t'\\ t&r'\end{array}\right)X^{-1},
\hspace{5mm}
X=\left(\begin{array}{cc} X_L^\pm&\\& X_R^\pm\end{array}\right).
\end{equation}

Here $r\,(r')$ describes left (right) to left (right) reflection, while $t\, (t')$ describes
left (right) to right (left) transmission ($t$ is not to be confused with time). 
These matrices have no time or Keldysh structure but still have
matrix structure in the space of left or right channel indices. The operators
$X_L^{\pm}(\tau)$ and $X_R^{\pm}(\tau)$ have diagonal Keldysh structure (denoted by the superscript $\pm$) and diagonal
time structure (here indicated by $\tau$ to avoid confusion with the transmission matrix $t$). 
They do not have internal channel structure and as a result the Keldysh action is
insensitive to the internal dynamics on the left- or right-hand sides. Our shorthand for the Keldysh scattering
matrix will be $XsX^{-1}$ where we remember that $s$ has no Keldysh structure.

We now consider the square of the generating functional $\mathcal Z$ and employ the first expression we obtained 
for it (Eq. \ref{firstform}) which retains Keldysh structure in the determinant.
\begin{equation}
\mathcal Z^2={\rm Det}\left[\frac{1+G}{2}+XsX^{-1}\frac{1-G}{2}\right]^2{\rm Det} s^\dagger
\end{equation}
Here we exploited the fact that $\hat s_-$ acts on half of Keldysh space together with the fact that $\hat s_+=\hat s_-$, i.e.
$s$ has no Keldysh structure, to write $\exp 2{\rm Tr}\,\ln \hat s_-={\rm Det}\,s$. We now shift $X$ to act on $G$ and define
\begin{equation}
\check G=X^{-1} G X\hspace{5mm}
P=\frac{1+\check G}{2}\hspace{5mm}
Q=\frac{1-\check G}{2}
\end{equation}
The operators $P$ and $Q$ are complementary projection operators i.e. $P^2=P$, $Q^2=Q$, $PQ=QP=0$ and $P+Q=I$.
Because of this, it holds that ${\rm Det}(P+sQ)={\rm Det}(P+Qs)$. Thus we find
\begin{equation}
\mathcal Z^{2}=e^{2\mathcal A}={\rm Det}(Ps^\dagger+sQ)
\end{equation}

The left channels are all connected to a single reservoir while the right channels are all connected to
a different reservoir. This means that the reservoir Green function has channel space structure
\begin{equation}
\check G=\left(\begin{array}{cc}\check G_L&\\&\check G_R\end{array}\right)
\end{equation}
where $G_L$ and $G_R$ have no further channel space structure.
At this point it is worth explicitly stating the structure of operators carefully. In general, an operator
carries Keldysh indices, indices corresponding to left and right, channel indices within the left or right sets of channels,
and time indices. However $P$, $Q$ and $s$ are diagonal or even structureless, i.e. proportional to identity in
some of these indices. Let us denote Keldysh indices with $k,k'\in\{+,-\}$, left and right with $\alpha,\alpha'\in\{L,R\}$,
channel indices within the left or right sets with $c,c' \in Z$ and time $t,t'\in R$. 
Then $P$ has the explicit form
\begin{equation}
P(k,k';\alpha,\alpha';c,c';t,t')=P(k,k';\alpha;t,t')\delta_{\alpha,\alpha'}\delta_{c,c'}.
\end{equation}  
The projection operator $Q$ has the same structure. The scattering matrix $s$ has the structure
\begin{equation}
s(k,k';\alpha,\alpha';c,c';t,t')=s(\alpha,\alpha';c,c')\delta_{k,k'}\delta(t-t').
\end{equation}

We now use the formula ${\rm Det}\left(\begin{array}{cc}A&B\\C&D\end{array}\right)={\rm Det}(A){\rm Det}(D-CA^{-1}B)$
to eliminate left-right structure from the determinant.
\begin{widetext}
\begin{eqnarray}
\mathcal Z^2&=&\left(\begin{array}{ll}P_L r^\dagger+ Q_L r& P_L t^\dagger+Q_Rt'\\
P_R\left.t'\right.^\dagger +Q_L t& P_R \left.r'\right.^\dagger+Q_Rr'\end{array}\right)\nonumber\\
&=&{\rm Det}\left(P_Lr^\dagger+Q_Lr\right){\rm Det}\left[P_R\left.r'\right.^\dagger+Q_Rr'-
\left(P_R\left.t'\right.^\dagger+Q_L t\right)\left(P_L \left.r^\dagger\right.^{-1}+Q_Lr^{-1}\right)
\left(P_Lt^\dagger+Q_r t'\right)\right]\nonumber\\
&=&\underbrace{{\rm Det}(P_Lr^\dagger+Q_Lr)}_a\nonumber\\
&&\times\underbrace{{\rm Det}\left[P_R(\left.r'\right.^\dagger-P_L\left.t'\right.^\dagger
\left.r^\dagger\right.^{-1}t^\dagger)+(r'-Q_Ltr^{-1}t')Q_R-P_R(
P_L\left.t'\right.^\dagger \left.r^\dagger\right.^{-1}t'
+Q_L\left.t'\right.^\dagger r^{-1}t')Q_R\right]}_b
\end{eqnarray}
\end{widetext}
Here it is important to recognize that the reflection and transmission matrices commute with the projection operators
$P_{L,R}$ and $Q_{L,R}$. Furthermore, notice that, in term $b$, the projection operator $P_R$ always appears on the
left of any product involving other projectors, while $Q_R$ always appears on the right. This means that in the 
basis where
\begin{equation}
P_R=\left(\begin{array}{cc}I&0\\0&0\end{array}\right)\hspace{3mm}
Q_R=\left(\begin{array}{cc}0&0\\0&I\end{array}\right)
\end{equation}
term $b$ is the determinant of an upper block-diagonal matrix. As such, it only depends on the diagonal blocks, so that 
the term $P_R(\ldots)Q_R$ may be omitted. Hence
\begin{align}
b={\rm Det}\Big[P_R(\left.r'\right.^\dagger&-P_L\left.t'\right.^\dagger
\left.r^\dagger\right.^{-1}t^\dagger)\nonumber\\&+(r'-Q_Ltr^{-1}t')Q_R\Big].
\end{align}

Now we invoke the so-called polar decomposition of the scattering matrix \cite{Bar94}
\begin{equation}
\begin{array}{ll}
r=u\sqrt{1-T}u'&\hspace{3mm}t'=iu\sqrt{T}v\\
t=iv'\sqrt{T}u'&\hspace{3mm}r'=v'\sqrt{1-T}v
\end{array}
\end{equation}
where $u$, $u'$, $v$ and $v'$ are unitary matrices and $T$ is a diagonal matrix with the transmission 
probabilities $T_n$ on the diagonal.
We evaluate term $a$ in the basis where $P_L$ and $Q_L$ are diagonal to find
\begin{eqnarray}
a&=&{\rm Det}\left(\begin{array}{cc}\left.u'\right.^\dagger\sqrt{1-T}u^\dagger&0\\0&u\sqrt{1-T}u'\end{array}\right)\nonumber\\
&=&{\rm Det}\left(I\sqrt{1-T}\right),
\end{eqnarray} 
Where $I=P_L+Q_L=P_R+Q_L$ is the identity operator $I(k,k';c,c';t,t')=\delta_{k,k'}\delta_{c,c'}\delta(t-t')$ in Keldysh, channel and 
time indices. For term $b$ we find
\begin{align}
b={\rm Det}\Big[P_R&\left(\sqrt{1-T}+P_L\frac{T}{\sqrt{1-T}}\right)\nonumber\\
&+\left(\sqrt{1-T}+Q_L\frac{T}{\sqrt{1-T}}\right)Q_R\Big]
\end{align}
Combining the expressions for $a$ and $b$ we find
\begin{equation}
\mathcal Z^2=e^{2\mathcal A}={\rm Det}\left[1-T(P_R Q_L+P_L Q_R)\right].
\end{equation}
Using the fact that $P_{L(R)}=(1+\check G_{L(R)})/2$ and $Q_{L(R)}=(1- \check G_{L(R)})/2$
and taking the logarithm we finally obtain the remarkable result
\begin{equation}
\mathcal A=\frac{1}{2}\sum_n{\rm Tr}\,{\rm ln}\left[1+\frac{T_n}{4}\left(\left\{\check G_L,\check G_R\right\}-2\right)\right]
\end{equation}
This formula was used in [\onlinecite{Kin03}] to study the effects on transport of electromagnetic interactions
among electrons. In [\onlinecite{Tob06}] the same formula was employed to study the output of a two-level measuring
device coupled to the radiation emitted by a QPC.
 
\section{Fermi Edge Singularity}
In this section we show how our formulas apply to a phenomenon known as the Fermi Edge Singularity.
The system under consideration is one of the most elementary examples of an interacting electron system.
The initial analysis  \cite{Mah67,Noz69} relied on diagrammatic techniques rather than the scattering approach or the Keldysh
technique, and was confined to equilibrium situations. Several decades later the problem was revisited in
the context of the scattering approach\cite{Aba04,Aba05}. An intuitive derivation of a determinant formula 
was given. Here we apply our approach to confirm the validity of this previous work. We find
exact agreement. This highlights the fact that the determinant formulation of the FES problem is also 
valid for multi-channel devices out of equilibrium, an issue not explicitly addressed in the existing 
literature.

The original problem \cite{Mah67,Noz69} was formulated for conduction electrons with a small effective 
mass and valence electrons with a large effective mass, bombarded
by x-rays. The x-rays knock one electron out of the valence band leaving behind an essentially stationary 
hole. Until the hole is refilled, it interacts through the coulomb interaction with the conduction electrons.
The x-ray absorption rate is studied. Abanin and Levitov reformulated the problem in the context of quantum transport
where an electron tunnels into or out of a small quantum dot that is side-coupled to a set of transport 
channels.

\begin{figure}[h]
\begin{center}
\includegraphics[width=.95 \linewidth]{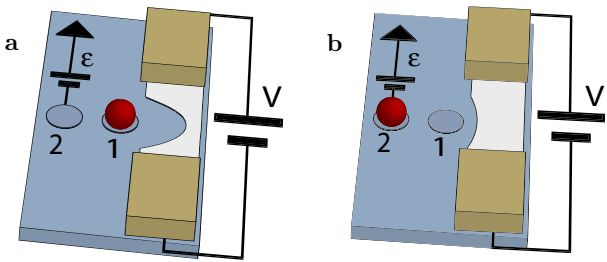}
\caption{A schematic picture of the system considered. It consists of a charge qubit
coupled to a QPC. The shape of the QPC constriction, and hence its scattering matrix,
depends on the state of the qubit. A gate voltage controls
the qubit level splitting $\varepsilon$. There is a small tunneling rate $\gamma$ between 
qubit states.\label{fig2}}
\end{center}
\end{figure}
We prefer to consider a slightly simpler setup that exhibits the same physics. The setup is illustrated in Fig. 
(\ref{fig2}). A Quantum Point Contact (QPC) interacts with a charge qubit. The shape of the QPC constriction depends on the 
state of the qubit.
The Hamiltonian for the system is
\begin{equation}
\mathcal H=\mathcal H_1\left|1\right>\left<1\right|+(\mathcal H_2+\varepsilon)\left|2\right>\left<2\right|
+\gamma(\left|1\right>\left<2\right|+\left|2\right>\left<1\right|)\label{Ham}
\end{equation}
The operators $\mathcal H_1$ ($\mathcal H_2$) describe the QPC electrons when the qubit is in 
state $\left|1\right>$, ($\left|2\right>$). They differ by a potential energy term, describing the
the pinching off of the QPC constriction depending on the state of the qubit.
We may take both Hamiltonians to be of the form (Eq. \ref{genham}) that we wrote down for a general scatterer.
The energy $\varepsilon$ is the qubit level splitting, an experimentally tunable parameter.
The QPC may or may not be driven by a voltage bias $V$.
  
QPC electrons do not interact directly with each other but rather with the qubit.  
This interaction is the only qubit relaxation mechanism included in our model.
We work in the limit $\gamma \to 0$ 
where the inelastic transition rates $\Gamma_{12,21}$
between qubit states are small
compared to the energies $eV$ and $\varepsilon$.
In this case, the qubit switching events can be regarded as independent and incoherent.

Now consider the qubit transition rate $\Gamma_{21}$, from state $\left|1\right>$ to $\left|2\right>$
as a function of the qubit level splitting $\varepsilon$.
To lowest order in the tunneling amplitude $\gamma$ it is given by
\begin{eqnarray}
\Gamma_{21}&=&2\gamma^2{\rm Re}\,\int_{-\infty}^0d\tau\,e^{i\varepsilon\tau}
\lim_{t_0\to-\infty}\exp\mathcal A(\tau)\nonumber\\
\exp\mathcal A(\tau)&=&{\rm tr}\left[e^{i\hat{H}_2\tau}e^{-i\hat{H}_1(\tau-t_0)}\rho_0 e^{-i\hat{H}_1t_0}\right].\label{rate}
\end{eqnarray}
This is the usual Fermi Golden Rule. The time $\tau$ over which we integrate can be interpreted as the time when
the qubit switches from $\left|1\right>$ to $\left|2\right>$.
The trace is over QPC states, and $\rho_0$ is the initial QPC density matrix.
We see that the expression for $\Gamma_{21}$ contains an instance of the Keldysh action $\mathcal A$ that
we have calculated. The correspondence requires us to set
\begin{eqnarray}
\mathcal H^+(t)&=&\mathcal H_1+(\mathcal H_2-\mathcal H_1) \theta(t-\tau)\theta(-t)\nonumber\\
\mathcal H^-(t)&=&\mathcal H_1.
\end{eqnarray}
In order to conform to the conventions of the existing literature, we write $\mathcal Z$ in the form
where the Keldysh structure has been removed (Eq. \ref{eqb}):
\begin{equation}
\mathcal A(\tau)={\rm tr\,ln}\,\left[\hat s_- (1-\hat f)+\hat s_+(\tau)\hat f\right]-{\rm tr\,ln}\,\hat s_-
\end{equation}
In this formula, $\hat s_-$ is the scattering matrix corresponding to $\mathcal H^-=\mathcal H_1$ when the qubit is in 
state $\left|1\right>$. It is proportional to identity in time-indices. The scattering matrix
$\hat s_+(\tau)$ corresponds to $\mathcal H^+$. It is still diagonal in time-indices but the diagonal elements 
$\hat s_+(\tau)_t$ are time-dependent. If we take the time it takes an electron to traverse the conductor 
to be much shorter than other time-scales such as the attempt rate of charge transfers, then 
\begin{equation}
\hat s_+(\tau)_t=\hat s_1 +(\hat s_2-\hat s_1)\theta(t-\tau)\theta(-t)
\end{equation}
where $\hat s_2$ is the scattering matrix associated with $\mathcal H_2$ when the qubit is in state $\left|2\right>$.
This expression first appeared in [\onlinecite{Aba04}].
In the language of the original diagrammatic treatment of the FES problem \cite{Mah67,Noz69}, it represents the
total closed loop contribution.

 We may also write this closed loop contribution as
\begin{equation}
e^{\mathcal A(\tau)}={\rm Det}\left[1+(\hat s_1^\dagger \hat s_2-1)\hat \Pi(\tau)\hat f\right]
\end{equation}
where $\hat \Pi$ is a diagonal operator in time-domain with a kernel that is a double step function
\begin{equation}
\Pi(\tau)_{t,t'}=\theta(-t)\theta(t-\tau).
\end{equation}
and the scattering matrices $\hat s_1$ and $\hat s_2$ no longer have time-structure.
We may work in the channel space basis where $\hat s_1^\dagger\hat s_2$ is diagonal. 
Its eigenvalues are $e^{i\lambda_k}$.
Suppose we are in zero-temperature equilibrium, then the filling factor $f$ is the same in every channel. In
the fourier transformed energy basis $f$ is simply a step function:
\begin{equation}
f_{\varepsilon,\varepsilon'}=\delta(\varepsilon-\varepsilon')\theta(-\varepsilon)
\end{equation}
Thus one finds
\begin{equation}
e^\mathcal A=\prod_{k}{\rm Det}\left[1+(e^{i\lambda_k}-1)\hat \Pi(\tau)\hat f\right].
\end{equation}
This determinant contains no channel structure any more. Operators only have one set of indices (time, or after
Fourier transform, energy). $\hat \Pi$ is a projection operator, diagonal in time-domain while $\hat f$ is a projection
operator in energy domain. Such a determinant is known as a Fredholm determinant.

The resulting transition rate is\cite{Mah67,Noz69,Aba04}
\begin{equation}
\Gamma_{21}(\varepsilon)=\theta(-\epsilon)\frac{1}{|\epsilon|}\left(\frac{|\varepsilon|}{E_{\rm c.o}}\right)^\alpha
\end{equation}
where $E_{\rm c.o}$ is a cut-off energy of the order of the Fermi energy measured from the bottom of the conduction band.
The exponent $\alpha$ is known as the orthogonality exponent. It may be calculated by evaluating the Fredholm determinant
analytically with Wiener-Hopf method. It is given in terms of the scattering matrices 
as\cite{Yam82,Aba04}
\begin{equation}
\alpha=\frac{1}{4\pi^2}\left|{\rm tr\,ln}^2\left(s_1^\dagger s_2\right)\right|
\end{equation}
with the trace being over channel indices. Inspired by the work of Abanin and Levitov \cite{Aba04,Aba05} we
considered the case where the QPC is driven by a voltage bias. The results of our study may be found in [\onlinecite{Sny07}].

\section{Conclusion}
In this paper we have derived several expressions for the Keldysh action $\mathcal A$ 
for a general multi-terminal, time-dependent scatterer. This object is defined as
the (logarithm of the) trace of the density matrix of the scatterer after evolution
forwards and backwards in time with different Hamiltonians:
\begin{align}
e^{\mathcal A}={\rm Tr}\Big[\mathcal T^+\exp&\left\{-i\int_{t_0}^{t_1}dt\;\mathcal H^+(t)\right\}\nonumber\\
&\rho_0\mathcal T^-\exp\left\{i\int_{t_0}^{t_1}dt\;\mathcal H^{-}(t)\right\}\Big].
\end{align}
Our main result is a compact formula for the 
action in terms of reservoir Green functions and the scattering matrix of the scatterer (Eq. \ref{eq0}). We have
shown how to perform the trace over Keldysh indices explicitly when reservoirs are characterized 
by filling factors. Thus we obtained a formula (Eq. \ref{eqa}) belonging to the same class as the Levitov-Lesovik
counting statistics formula. We have also explicitly performed the trace over channel indices
for a two terminal scatterer (Eq. \ref{eqb}). In this case we demonstrated that the Keldysh action only depends
on the scattering matrix through the eigenvalues of the transmission matrix. To illustrate the
utility of the Keldysh action, and confirm the correctness of our results, we considered 
Full Counting statistics and the Fermi Edge singularity. We found that our results agree with the
existing literature.

\end{document}